\documentclass[11pt]{article}
\usepackage{authblk}
\usepackage{fullpage}
\usepackage{amssymb,amsmath}
\usepackage[T1]{fontenc}
\usepackage{siunitx}
\usepackage[version=3]{mhchem}
\usepackage{xr}
\usepackage{bm}
\usepackage{pdfpages}
\usepackage[a4paper, total={6in, 10in}]{geometry}
\usepackage{graphicx}% Include figure files
\usepackage{dcolumn}% Align table columns on decimal point
\usepackage{bm}% bold math
\usepackage{color}
\usepackage{multirow}
\usepackage{tikz}
\usepackage[font=small,labelfont=bf, center]{caption}
\usepackage{setspace}
\usepackage[title]{appendix}

\usepackage{float}
\usepackage{braket}
\usepackage{soul}
\usepackage[unicode=true]{hyperref}
\hypersetup{colorlinks=true,
            citecolor=blue,
            linkcolor=blue}

\usepackage[authoryear,sort&compress]{natbib}
\usepackage{setspace}

\floatname{algorithm}{Algoritmo}

\newcommand{\bo}{\raise-1mm\hbox{\Large$\Box$}}

\newcommand{\s}[1]{\mbox{\tiny{#1}}}
\begin{document}

%\title{How movement toward attractive regions determines population spread and critical habitat size in heterogeneous landscapes}
%\title{Population spread and critical habitat size in a heterogeneous landscape with space-dependent advection drift}
%\title{How space-dependent advection drift determines population spread and critical habitat size in heterogeneous landscapes}
\title{Movement bias in asymmetric landscapes and its impact on population distribution and critical habitat size}

\author[1,2]{Vivian Dornelas\footnote{These authors contributed equally to this work and share first authorship}}
\author[1]{Pablo de Castro$^*$}
\author[3,4,5]{Justin M. Calabrese}
\author[5]{William F. Fagan}
\author[3,1]{Ricardo Martinez-Garcia\footnote{Correspondence: r.martinez-garcia@hzdr.de}}

\vspace{1cm}
\affil[1]{\footnotesize ICTP -- South American Institute for Fundamental Research and Instituto de F\'isica Te\'orica, Universidade Estadual Paulista -- UNESP, S\~ao Paulo, Brazil.}
\affil[2]{ National Institute of Chemical Physics and Biophysics - Akadeemia Tee 23, Tallinn 12618, Estonia.}
\affil[3]{Center for Advanced Systems Understanding (CASUS); Helmholtz--Zentrum Dresden--Rossendorf (HZDR), Görlitz, Germany.}
\affil[4]{Department of Ecological Modelling, Helmholtz Centre for Environmental Research -- UFZ, Leipzig, Germany.}
\affil[5]{Department of Biology, University of Maryland, College Park, MD, USA.}
\normalsize

\date{}
\maketitle

\begin{abstract}
Ecologists have long investigated how demographic and movement parameters determine the spatial distribution and critical habitat size of a population. However, most models oversimplify movement behavior, neglecting how landscape heterogeneity influences individual movement. We relax this assumption and introduce a reaction-advection-diffusion equation that describes population dynamics when individuals exhibit space-dependent movement bias toward preferred regions. Our model incorporates two types of these preferred regions: a high-quality habitat patch, termed `habitat', which is included to model avoidance of degraded habitats like deforested regions; and a preferred location, such as a chemoattractant source or a watering hole, that we allow to be asymmetrically located with respect to habitat edges. In this scenario, the critical habitat size depends on both the relative position of the preferred location and the movement bias intensities. When preferred locations are near habitat edges, the critical habitat size can decrease when diffusion increases, a phenomenon called the drift paradox. Also, ecological traps arise when the habitat overcrowds due to excessive attractiveness or the preferred location is near a low-quality region. Our results highlight the importance of species-specific movement behavior and habitat preference as drivers of population dynamics in fragmented landscapes and, therefore, in the design of protected areas.
\end{abstract}

\maketitle

%\tableofcontents
\newpage
\section{Introduction}
\label{intro}

Habitat destruction and fragmentation result in smaller and more isolated suitable habitat patches where extinctions are more likely to occur \citep{turner1996species,Lord1999, Franklin2002,ferraz2003rates, Nauta2022}. The viability of a population in each of these patches depends on the balance between growth inside the patch and population losses, mainly due to dispersal through habitat edges. The interplay between these two processes determines the spatial pattern of population density within a patch and defines the minimum area required to sustain the population. This patch-size threshold is often termed the critical patch size or critical habitat size \citep{kierstead1953size}. Thus, understanding the interaction between demographic and movement processes is key to determining critical patch sizes across species \citep{pereira2006}, which has important implications for conservation, such as in the design of protected areas or ecological corridors \citep{Cantrell1999,Ibagon2022}. Moreover, determining the expected spatial pattern of population density in patches larger than the critical size can improve understanding of population responses to further habitat destruction.

Much of our current understanding of how demographic and movement processes determine population spatial patterns and survival in habitat patches comes from mathematical models. The most common models to determine critical habitat sizes consist of a reaction-diffusion equation describing the spatio-temporal dynamics of a population density in a bounded region. Within this family of models, the simplest ones assume purely diffusive dispersal coupled with exponential growth and are commonly called KISS models \citep{Skellam1951,kierstead1953size}. Due to these highly simplified assumptions, KISS models lead to analytical expressions for the critical patch size. Some of these predictions have been confirmed in microcosm experiments with microbial populations \citep{Lin2004, Perry2005}.
%and in larger-scale manipulative experiments and field observations \citep{turner1996species,ferraz2003rates,pereira2006}. 

Departing from KISS models, researchers have developed refined models with more realistic movement descriptions. Some of these extensions include space-dependent diffusion within the patch \citep{dos2020critical, colombo2018nonlinear}, responses to habitat edges \citep{Fagan1999,maciel2013individual,cronin2019effects}, and various sources of non-random movement, such as a constant external flow \citep{Pachepsky2005,ryabov2008population, Vergni2012, speirs2001population} or a chemoattractant secreted by the population \citep{Kenkre2008}. Other extensions of the KISS models have explored more complex growth dynamics, such as Allee effects \citep{Alharbi2016,cronin2020modeling},  time-varying environments \citep{zhou2017discrete}, or heterogeneity in population growth, either through time-dependent demographic rates \citep{Ballard2004,Colombo2016} or by introducing a finer spatial structure of habitat quality within the patch \citep{Cantrell2001,maciel2013individual,fagan2009interspecific}. Finally, a few studies have considered an environmental gradient that results in spatially-dependent demographic rates  and induces migration toward higher quality regions within the patch and obtained the critical patch size \citep{Cantrell1991,Cantrell2001,cantrell2006movement} and spatial population distributions depending on the type of boundary conditions \citep{belgacem1995effects}.

Despite these efforts to refine classical KISS models, some movement features routinely present in real populations remain underexplored or have been investigated under very specific assumptions. For example, individuals often show a tendency to move toward certain habitat regions where they concentrate, which makes population ranges smaller than the total amount of habitat available \citep{van2016movement,kapfer2010modeling}. While a considerable effort has focused on understanding how and why individuals show these patterns of space use \citep{Nathan2008,Jeltsch2013,fleming2014fine}, their population-level consequences, especially in fragmented landscapes, have been less explored. Motivated by colonial central-place foragers such as ants, beavers, and colonial seabirds, one particular study obtained, numerically, the critical patch size when the home-range center is the same for all the individuals and is located at the center of the habitat patch (i.e., at the same distance from both habitat edges; \citeauthor{fagan2007population} \citeyear{fagan2007population}). At a larger scale, theoretical studies have also derived some results regarding the impact of movement bias toward the center of good-quality patches for invasive spread in periodic patchy environments \citep{kawasaki2012impact}. All these studies, however, consider that movement is biased toward the center of good-quality patches. This might be a realistic assumption for range-resident movement after individuals adapt their home range following habitat degradation events. In many other cases, however, animals are attracted to regions that can be arbitrarily distributed within the available habitat patch (e.g., water resources that are located close to one of the habitat edges).

Here, we further extend classical KISS models to account for generic space-dependent deterministic movement in which attraction regions are arbitrarily located within a habitat patch. Our motivation is to study scenarios with movement bias toward preferred points in space such as a chemoattractant source \citep{tyson2007modelling} or the location of a special resource, like a watering hole or a salt-lick. %Regarding the latter example, there has been strong evidence that impalas stay closer to water sources during the dry season \citep{hilbers2015modeling,homolavc2021importance}. In all these cases, these preferred locations will, in general, not be located at the center of the good-quality habitat.}
We study how this additional movement bias influences spatial population distribution in a heterogeneous landscape and the critical habitat patch size that ensures population survival. We consider the simple one-dimensional scenario with a finite high-quality habitat patch embedded in a low-quality ``matrix'' (characterized, for instance, by deforestation or desertification) with high mortality. Using both numerical and analytical methods, we measure critical patch size and spatial patterns of population density for different matrix mortality levels. We also vary the intensity of two deterministic space-dependent movement components relative to random dispersal: a bias to preferred landscape locations and avoidance of degraded habitats. We find that the total population lost due to habitat degradation and the critical patch size depend nonlinearly on both key movement parameters and on the spatial distribution of habitat. This modeling approach provides a general framework to investigate how habitat selection within a fragmented landscape determines both population spatial distributions and critical patch size. Our results emphasize the importance of incorporating covariation between movement behavior and landscape features when investigating population dynamics in heterogeneous landscapes.

%=================================================
\section{Material and Methods}\label{sec:mat-met}

\subsection{Model formulation}\label{subsec:model}

\begin{figure}[!ht]
    \centering
    \includegraphics[width=0.5\columnwidth]{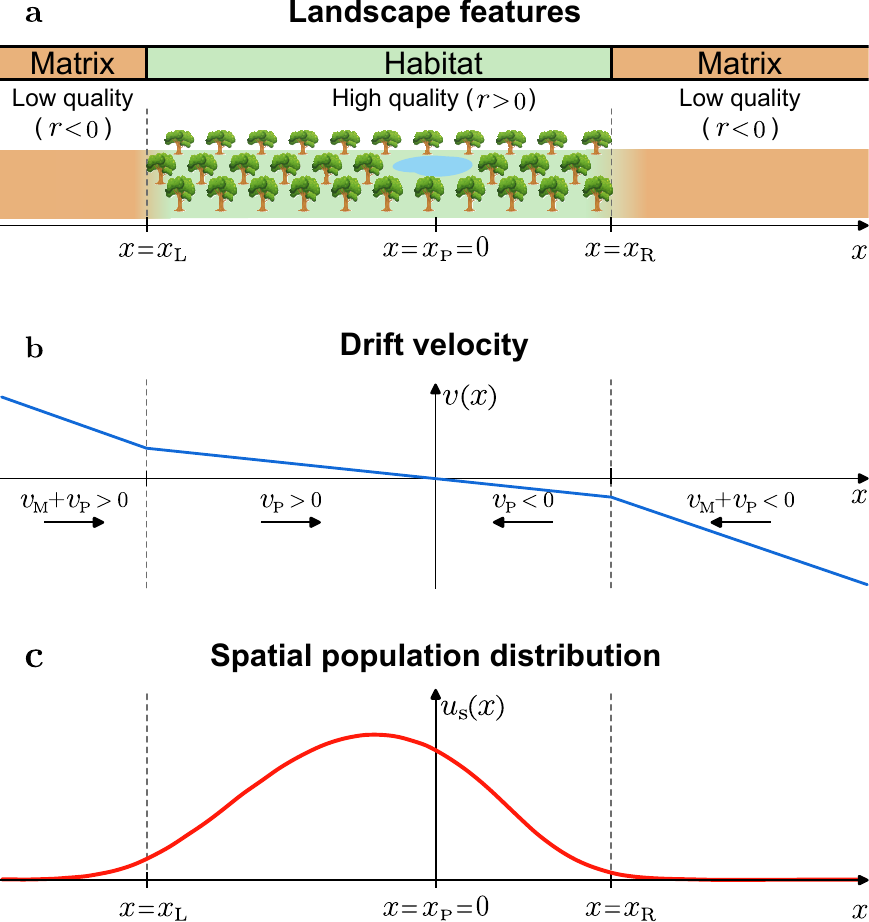}
    \caption{Model schematics. (a) Landscape features, showing a high-quality habitat ($r>0$) and an attractive location (represented as a watering hole at position $x=x_{\s{P}}=0$) surrounded by low-quality matrix regions ($r<0$). Habitat edges are located at $x=x_{\s{L}}$ and $x=x_{\s{R}}$. In this example, $x_{\s{R}}$ is positive. The total habitat size is $L=|x_{\s{R}}-x_{\s{L}}|$. (b) Spatial dependence of the drift velocity (deterministic movement component). Velocity $v_{\s{P}}$ points to the attractive location [Eq.\,\eqref{eq:veldef}]. For individuals in the matrix, an additional term ($v_{\s{M}}$) points to the edges [Eq.\,\eqref{eq:veldefmatrix}]. (c) Emerging stationary population density distribution $u_{\s{s}}(x)$, peaking to the left of $x=0$.}
    \label{fig:scheme}
\end{figure}

We consider a one-dimensional heterogeneous landscape with a habitat patch embedded in an infinite matrix (see Fig.\,\ref{fig:scheme}a). The left and right habitat patch edges are located at $x=x_{\s{L}}$ and $x=x_{\s{R}}$, respectively, and the habitat patch size is $L=|x_{\s{R}}-x_{\s{L}}|$. The landscape is occupied by a single-species population, which we describe via a continuous density field $u(x,t)$. This population density changes in space and time due to demographic processes and dispersal. For the birth/death dynamics, we assume that the population follows logistic growth with net reproduction rate $r$ and intraspecific competition intensity $\gamma$. The net growth rate is constant within each type of region but different between regions: $r(x)=r_{\s{H}}>0$ inside the habitat patch (high-quality, low-mortality region) and $r(x)=r_{\s{M}}<0$ in the matrix (low-quality, high-mortality region). The matrix mortality rate $r_{\s{M}}$ defines the degree of habitat degradation, with the limit $r_{\s{M}}\rightarrow-\infty$ representing complete habitat destruction. For finite mortality rates, whether an individual dies in the matrix or not is determined by the mortality rate itself and the time the individual spends in the matrix. Therefore, when the matrix is not immediately lethal, the population density outside the habitat patch is not zero.
For dispersal, we consider two different movement components: random dispersal with constant diffusion coefficient $D$, and a deterministic tendency of individuals to move toward attractive regions with space-dependent velocity $v(x)$, therefore accounting for the effect of landscape heterogeneity in movement behavior. Importantly, this attractive term in our model generates movement bias toward regions that are not necessarily of higher habitat quality. The actual velocity of an individual is thus equal to $v(x)$ plus a stochastic contribution that comes from diffusion. Combining these demographic and movement processes, the dynamics of the population density is given by
\begin{equation}
\label{eq:PDE-popdens}
\frac{\partial u(x,t)}{\partial t} =r(x) u(x,t)-\gamma u(x,t)^2+ D\frac{\partial^2 u(x,t)}{\partial x^2} - \frac{\partial}{\partial x}\bigg(v(x)u(x,t)\bigg).
\end{equation}
The functional form of the advection velocity $v(x)$ depends on landscape features, with attractive locations corresponding to $x$ coordinates with slower velocity. We consider two different types of attractive regions. 

First, we incorporate a tendency to move toward an attractive \textit{location} (such as a chemoattractant source or a watering hole) with velocity $v_{\s{P}}(x)$. We choose 
\begin{equation}\label{eq:veldef}
    v_{\s{P}}(x)=-\tau_{\s{P}}^{-1}(x-x_{\s{P}}),
\end{equation}
where we assumed that the velocity at which individuals tend to move toward attractive landscape regions increases linearly with the distance to the focus of attraction. This is similar to how simple data-driven models for range-resident movement implement attraction to home-range center at the individual level \citep{Dunn1977,Noonan2019}. The prefactor $\tau_{\s{P}}^{-1}$ is the attraction rate toward the attractive location and defines the typical time that individuals take to re-visit $x_{\s{P}}$. In the following, we use $x_{\s{P}}=0$ in all our calculations, such that the locations of the habitat edges are measured relative to the focus of attraction.

Second, we consider that individuals in the matrix tend to return to the habitat patch with velocity $v_{\s{M}}(x)$, and therefore, we incorporate an additional attraction term biasing movement from the matrix toward its closest habitat edge. Again, we consider a linear spatial dependence, but now only for individuals in the matrix:
\begin{equation}\label{eq:veldefmatrix}
v_{\s{M}}(x) = \begin{cases} 
      -\tau^{-1}_{\s{M}}(x-x_{\s{L}}), & x<x_{\s{L}}  \\
      0, & x_{\s{L}}\leq x\leq x_{\s{R}}\,. \\
      -\tau^{-1}_{\s{M}}(x-x_{\s{R}}), & x>x_{\s{R}}
   \end{cases}
   \end{equation}
The prefactor $\tau^{-1}_{\s{M}}$ is the edge attraction rate that modulates the strength of the matrix-to-habitat attraction $v_{\s{M}}(x)$. In the habitat, $v_{\s{M}}(x)=0$, whereas in the matrix, it is equal to $\tau_{\s{M}}^{-1}$ multiplied by the distance to the closest edge. Moreover, the velocity $v_{\s{M}}(x)$ always points toward the habitat patch, therefore biasing the movement of the individuals in the matrix toward the habitat-matrix edges. This matrix avoidance drift assumes that individuals remain aware of the direction in which the favorable habitat is located, which extends previous models for movement response to habitat edges that act only at the habitat-matrix boundary \citep{Fagan1999,maciel2013individual,cronin2019effects}. Putting together the movement toward the attractive location and the matrix avoidance bias, we obtain a velocity of the form $v(x) = v_{\s{P}}(x)+v_{\s{M}}(x)$ (Fig.\,\ref{fig:scheme}b). Because $v(x)$ increases linearly with the distance from the attractive location, the population does not spread indefinitely even in infinite habitat patches, which leads to finite population sizes in this limit case (Fig.\,\ref{fig:scheme}c). Also, because the width of the population spatial distribution is finite, the organisms never experience an infinite bias to the attractive location.
%for $v(x)$ and Fig.\,\ref{fig:scheme}c for the spatial population distribution emerging from it). We highlight that, for an infinite habitat patch, the total population is not infinite since attraction concentrates individuals around the attractive location, leading to a limited total population through competition. As a result, the population goes to zero at some finite distance from the attractive location, and therefore the maximum drift velocity $v(x)$ that emerges in the system is finite. 
We provide a summary of the model parameters in Table \ref{tab1}.
\begin{table}[!h]
\centering
\footnotesize
\begin{tabular}{ |c|c|c|  }
 \hline
 Sym. & Parameter & Dimensions \\
 \hline
$r_{\s{H}}$   & Habitat net growth rate  &time$^{-1}$\\
$r_{\s{M}}$   & Matrix net rate  &time$^{-1}$ \\ %  $-\infty$ or $\left[-10^{-3},-10^{1}\right]$\\
$\gamma$ & Intensity of intraspecific compet. & (time\,$\times$\,density)$^{-1}$ %1.0 
\\ $D$   & Diffusion coefficient & space$^2$/time\\
$x_{\s{P}}$   & Center attractive location & space\\
$x_{\s{L}}$   & Position of left habitat edge& space \\ %$\left[-1.5,7\right]$\\
$x_{\s{R}}$   & Position of right habitat edge & space \\ %$\left[-1.5,7\right]$\\
$\tau_{\s{P}}^{-1}$   & Attraction rate to attractive location & time$^{-1}$ \\ %$\left[0.006,1.5\right]$\\
$\tau_{\s{M}}^{-1}$   & Matrix-to-habitat attraction rate & time$^{-1}$ \\% $\left[10^{-3},10^{1}\right]$\\
 \hline
 \end{tabular}
\caption{\label{tab1} Summary of model parameters with symbols and dimensions. Units are arbitrary. Specific values are provided in figure captions.}
\end{table}

\subsection{Model analysis}

We analyze the stationary solutions of Eq.\,\eqref{eq:PDE-popdens} using a combination of a semi-analytical method on the linearized model equation and numerical simulations of the full nonlinear equation. We use both approaches in the $r_{\s{M}}\rightarrow-\infty$ limit and perform only numerical simulations in the more general case with finite $r_{\s{M}}$.

\subsubsection{Semi-analytical determination of critical habitat for $r_{\s{M}}\rightarrow-\infty$} \label{subsec:lsa}

In the $r_{\s{M}}$$\rightarrow$$-\infty$ limit, individuals die instantaneously upon reaching the matrix, and we can replace the dynamics of the population density in the matrix by absorbing boundary conditions at the habitat edges, $u(x_{\s{L}},t)=u(x_{\s{R}},t)=0$. In this regime, the movement component that attracts individuals to the habitat edge, $v_{\s{M}}$, has no effect on the dynamics. To determine the critical locations of the habitat edge $(x_{\s{L}}, x_{\s{R}})$ that lead to population extinction for a given set of movement parameters, we analyze the steady state of the linearized version of Eq.\,\eqref{eq:PDE-popdens} and obtain the conditions under which the extinction state $u(x,t\rightarrow\infty)\equiv u_s(x)=0$ is the only possible solution. Such linearization is possible because the population loss is continuous and the transition to an extinction state occurs from a small population size \citep{okubo1972note}. To perform this analysis, we neglect the quadratic term in the logistic growth and take the limit $t\rightarrow\infty$ in Eq.\,\eqref{eq:PDE-popdens}, which is equivalent to setting $\partial_t u(x,t)=0$. In this limit, Eq.\,\eqref{eq:PDE-popdens} becomes an ordinary differential equation that we solved using a symbolic calculation software. This solution is
\begin{equation} \label{eq:solution}
u_{\s{s}}(x)=\exp\left(-\frac{x^2}{2 \tau_{\s{P}}D}\right)\left[a
H_{r\tau_{\s{P}}}\left(\frac{x}{\sqrt{2D\tau_{\s{P}}}}
	\right)+\,b \,\, _1F_1\left(-\frac{r \tau_{\s{P}}}{2};\frac{1}{2};\frac{x^2}{2
	D\tau_{\s{P}}}\right)\right],
\end{equation}
where $a$ and $b$ are constants that depend on the boundary conditions, $_1F_1$ is the confluent hypergeometric function of the first kind, and $H_n(x)$ is the Hermite polynomial, with $n$ being a real, not necessarily integer, number \citep{arfken1999mathematical}. 

Imposing absorbing boundary conditions at the habitat edges on Eq.\,\eqref{eq:solution}, 
\begin{eqnarray}\label{eq:syssol}
   u_{\s{s}}(x_{\s{R}}) &=& 0 \\
   u_{\s{s}}(x_{\s{L}}) &=& 0
\end{eqnarray}
we obtain a system of two equations for $a$ and $b$ that can be used to determine whether the solution $u_s(x)=0$ is the only one possible. For this system of equations Eq.\,\eqref{eq:syssol} to have non-trivial solutions (that is, different from $a=b=0$), its determinant has to be zero. With this condition for the determinant and assuming that $x_{\s{L}}$ is fixed, we obtain a transcendental equation in $x_{\s{R}}$ that we can solve numerically to obtain the critical location of the right habitat edge, $x_{\s{R,C}}$. The numerical solution to this transcendental equation is obtained in the following way. First, we take the case whose solution we know analytically ($\tau_{\s{P}}^{-1}=0$). Then, we increase $\tau_{\s{P}}^{-1}$ by small increments and use as an initial guess the solution to the previous $\tau_{\s{P}}^{-1}$ considered.

\subsubsection{Numerical solution of the nonlinear model equation}\label{subsec:num}
%We perform all numerical simulations using a standard Euler scheme that is central in space and forward in time. The initial condition for $u$ is a small positive number plus random spatial fluctuations homogeneously distributed with zero mean. In the $r_{\s{M}}\rightarrow-\infty$ limit, we further ensure that the initial condition obeys the absorbing boundary conditions at the habitat edges. We integrate Eq.\,\eqref{eq:PDE-popdens} for a variety of habitat patch sizes keeping $x_{\s{L}}$ constant and decreasing $x_{\s{R}}$ systematically until the population reaches a stationary extinction state. Using this procedure, we can calculate $x_{\s{R,C}}$ and hence the critical patch size defined as the habitat patch size at which the steady-state total population transitions from non-zero to zero (see Fig.\,\ref{fig:sppatt} for spatial patterns of population density as $x_{\s{R}}$ decreases, with $x_{\s{R}}>x_{\s{R,C}}$). 

We performed all numerical simulations using a standard Euler scheme that is central in space and forward in time \citep{press2007numerical,github} and run them until the population spatial profile was constant up to a precision of $10^{-20}$. We used an initial condition for the density, $u(x,0)$, consisting of a small random spatial fluctuation uniformly distributed between $0$ to $10^{-4}$, but our results do not depend on the choice of the initial condition, as expected for this type of nonlinear dynamical system. To simulate an edge placed at ``infinity'', we implemented absorbing boundary conditions at a position far enough from the origin, ensuring that the population density there is smaller than $10^{-20}$ in the steady state. Imposing this condition is possible even for infinite systems because movement bias toward the attractive location prevents the population from spreading indefinitely (Fig.\,\ref{fig:scheme}c). In the limit case of $r_{\s{M}}\rightarrow-\infty$, we simulated the model equation \eqref{eq:PDE-popdens} in a finite domain $[x_{\s{L}},x_{\s{R}}]$ with absorbing boundary conditions at both habitat edges.

Using this numerical setup, we calculated $x_{\s{R,C}}$ integrating Eq.\,\eqref{eq:PDE-popdens} for a variety of habitat patch sizes, keeping $x_{\s{L}}$ constant and far enough from the center as described above. Then, we systematically reduced $x_{\s{R}}$ until the population underwent a transition from survival to extinction (see Fig.\,\ref{fig:sppatt} for spatial patterns of population density as $x_{\s{R}}$ decreases, with $x_{\s{R}}>x_{\s{R,C}})$. Finally, we also use the numerical solutions of Eq.\,\eqref{eq:PDE-popdens} to measure population loss due to habitat degradation. For this purpose, we introduce a dimensionless quantity, $\eta$, defined as the total population size sustained by a finite habitat patch of size $L$ divided by the total population size sustained by an infinite habitat patch. Such \textit{remaining population fraction} is thus
\begin{equation}\label{eq:eta}
    \eta\equiv \frac {N_{\s{T}}}{N_{\s{T}}^{\infty}},
\end{equation}
where $N_{\s{T}}$ and $N_{\s{T}}^{\infty}$ are the total population sizes for finite and infinite habitat patches, respectively. We obtain these population sizes by integrating the population density over the entire landscape, including the matrix.

%=================================================
\section{Results}\label{sec:results}

\subsection{Perfectly absorbing matrix: the $r_{\s{M}}\rightarrow-\infty$ limit}\label{subsec:rminf}

\begin{figure}[!h]
	\centering
	\includegraphics[width=0.6\columnwidth]{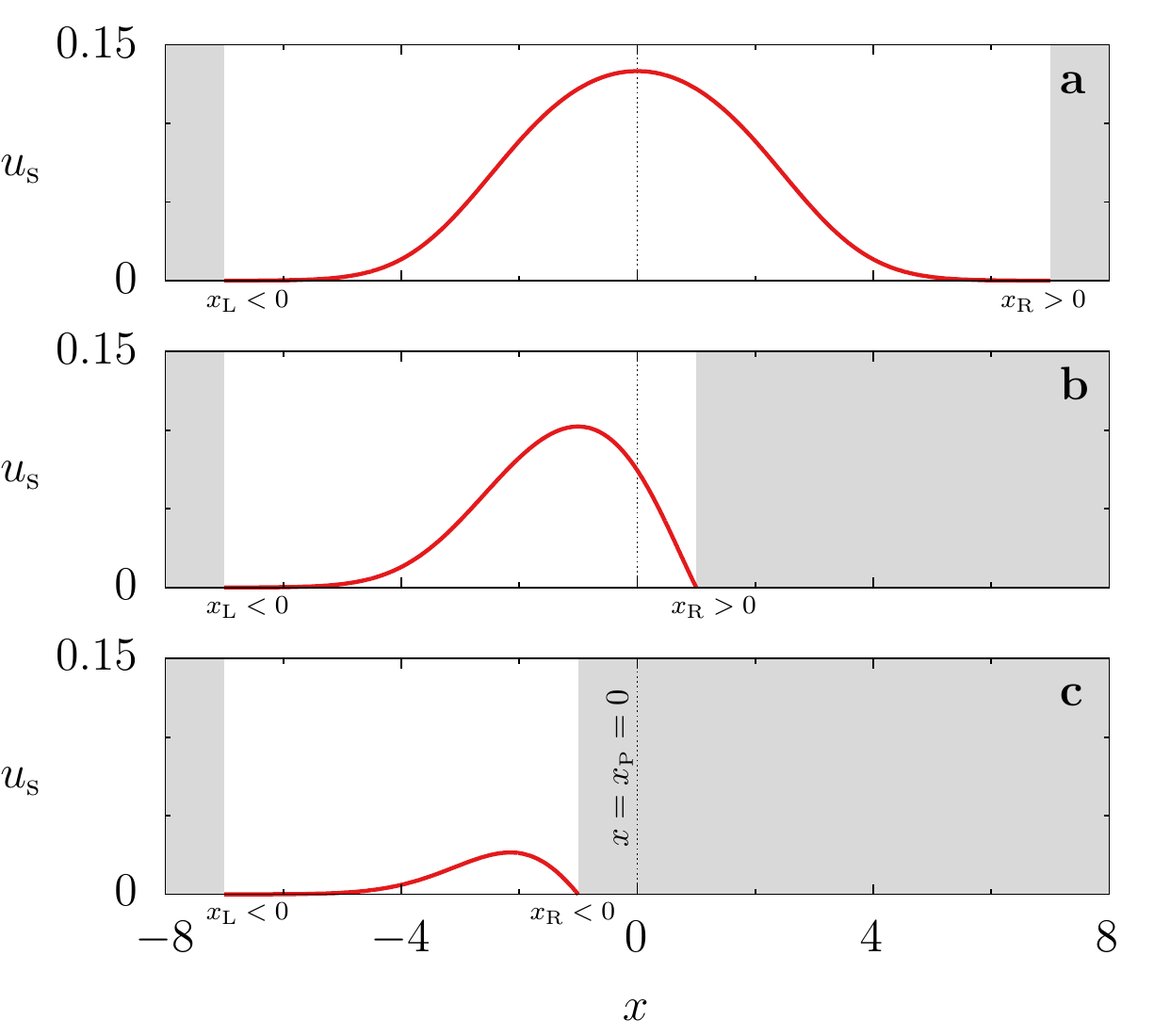}
	\caption{Stationary population density distribution $u_{\s{s}}(x)$ obtained from a numerical integration of Eq.\,\eqref{eq:PDE-popdens}. Space-dependence movement bias $v_{\s{P}}(x)$ is active and draws the population toward the attractive location $x=x_{\s{P}}=0$. (a) $x_{\s{R}}=7$ and $x_{\s{L}}=-7$, (b) $x_{\s{R}}=1$ and $x_{\s{L}}=-7$, and (c) $x_{\s{R}}=-1$ and $x_{\s{L}}=-7$. Other parameters: $r_{\s{H}}=0.1$, $\gamma=1$, $D=0.1$, $x_{\s{P}}=0$, and $\tau_{\s{P}}^{-1}=0.05$. Also, $r_{\s{M}}\rightarrow-\infty$. Gray regions represent the matrix.}
	\label{fig:sppatt}
\end{figure}

We first consider the simplest scenario in which individuals die instantaneously after they reach the habitat edges. In this limit, the population density is always zero in the matrix and, therefore, the movement component that biases individuals in the matrix toward the habitat edges is irrelevant. Movement is thus solely driven by random diffusion and the bias toward the attractive location $x=x_{\s{P}}=0$. 
In large habitat patches, space-dependent movement leads to the accumulation of population density very close to regions with slower movement. However, as the habitat patch decreases in size and regions with slower movement get closer to one of the habitat edges, the spatial pattern of population density changes due to mortality at the habitat edge and the maximum of population density shifts further away from the attractive location and towards the patch center (Fig.\,\ref{fig:sppatt}). This asymmetric pattern of space occupation due to space-dependent movement contrasts with well-known results for purely diffusive movement, for which population density reaches its maximum in the center of the habitat patch \citep{Holmes1994}, and significantly alters population loss due to habitat degradation.

\begin{figure*}[!h]
	\centering
	\includegraphics[width=0.9\textwidth]{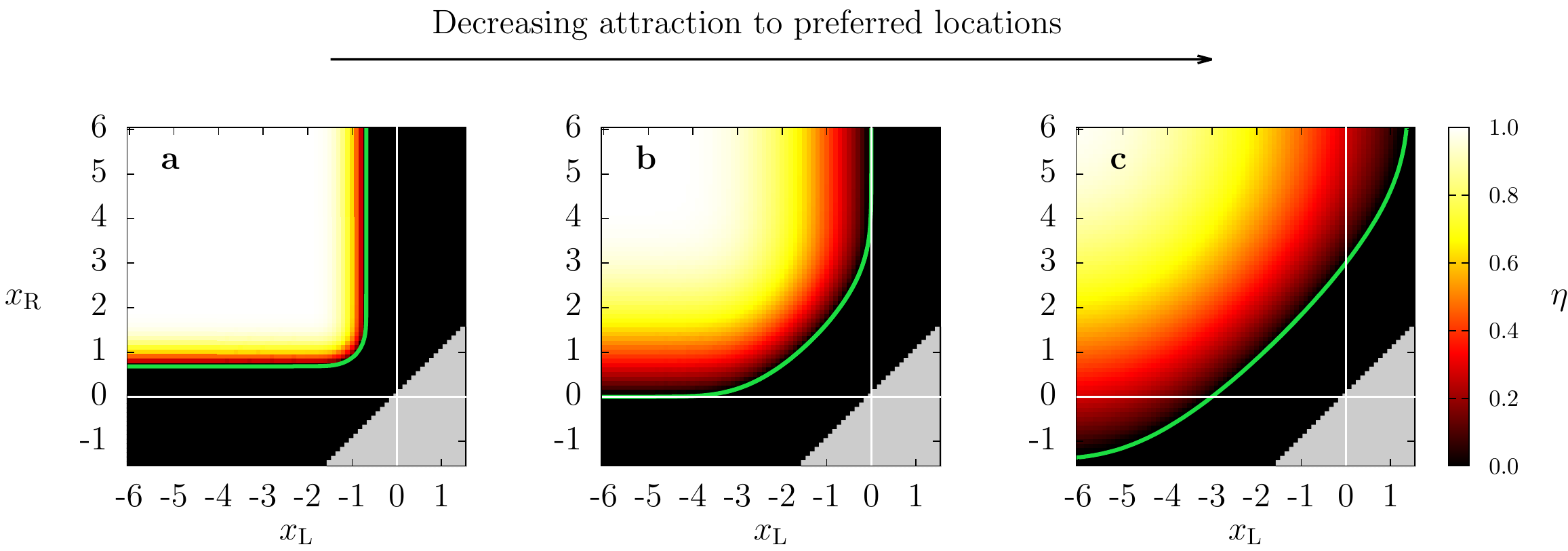}
	\caption{Remaining population fraction, $\eta$, for different habitat configurations $(x_{\s{L}},x_{\s{R}})$. (a) $\tau_{\s{P}}^{-1}=0.5$, (b) $\tau_{\s{P}}^{-1}=0.1$, and (c) $\tau_{\s{P}}^{-1}=0.05$. Other parameters: $r_{\s{H}}=0.1 $, $D=0.1$, and $\gamma=1$. Also, $r_{\s{M}}\rightarrow-\infty$. The solid green line shows the critical patch size obtained from the semi-analytical method while the color map comes from numerical simulations.}
	\label{fig:SurvivalRegion}
\end{figure*}

First, to understand population loss due to habitat degradation, we use the remaining population fraction, $\eta$, defined in Eq.\,\eqref{eq:eta}. This remaining population fraction is maximal when the attractive location is at the same distance from the two habitat edges, and it decays symmetrically about the line $x_{\s{R}} = -x_{\s{L}}$. Moreover, this decay is sharper for stronger bias toward the attractive location (Fig.\,\ref{fig:SurvivalRegion} and \,\ref{fig:SM1}). Finally, when the distance between the attractive location and one of the habitat edges is sufficiently large, further increasing the habitat size does not change the remaining population fraction because population loss through habitat edges is negligible, except for $\tau_{\s{P}}^{-1} = 0$.

Regarding the critical patch size, when the bias to the attractive location is strong, represented by higher values of $\tau_{\s{P}}^{-1}$, the population is localized around the attractive location $x_{\s{P}}=0$, and it goes extinct when the attractive location is within the habitat patch but close to one of its edges (Fig.\,\ref{fig:SurvivalRegion}a, b). When the bias to the attractive location decreases, however, the population can survive even if the attractive location is outside the habitat and the mortality in the matrix is infinite (Fig.\,\ref{fig:sppatt}c and \ref{fig:SurvivalRegion}c). This scenario would correspond to a situation where habitat destruction places the attractive location outside the habitat and individuals have not adapted their movement behavior to this landscape modification. As a result, individuals preferentially move toward regions with low habitat quality, which can be understood as an example of an ecological trap \citep{robertson2006framework,weldon2005effects,lamb2017forbidden}. 

To further investigate how the distance between the attractive location and the habitat edges determines the critical patch size for different movement parameters, we calculate the critical location of the right habitat edge $x_{\s{R,C}}$ assuming that the left habitat edge is fixed and far from the attractive location. In these conditions, if $x_{\s{R}}$ is also large, mortality through the left edge is negligible, but it becomes significant for smaller habitat patches. This setup mimics a situation where an initially large patch shrinks due to continued habitat destruction, slowly enough that the population distribution is at equilibrium for each particular habitat configuration, until it reaches a critical size and the population collapses. We find that the critical patch size is a nontrivial function of the intensity of the movement bias toward the attractive location and the distance between habitat edges and the attractive location. When $\tau_{\s{P}}^{-1}=0.1$, $x_{\s{R,C}}=0$ regardless of the value of $D$.

\begin{figure}[H]
	\centering
	\includegraphics[width=0.5\columnwidth]{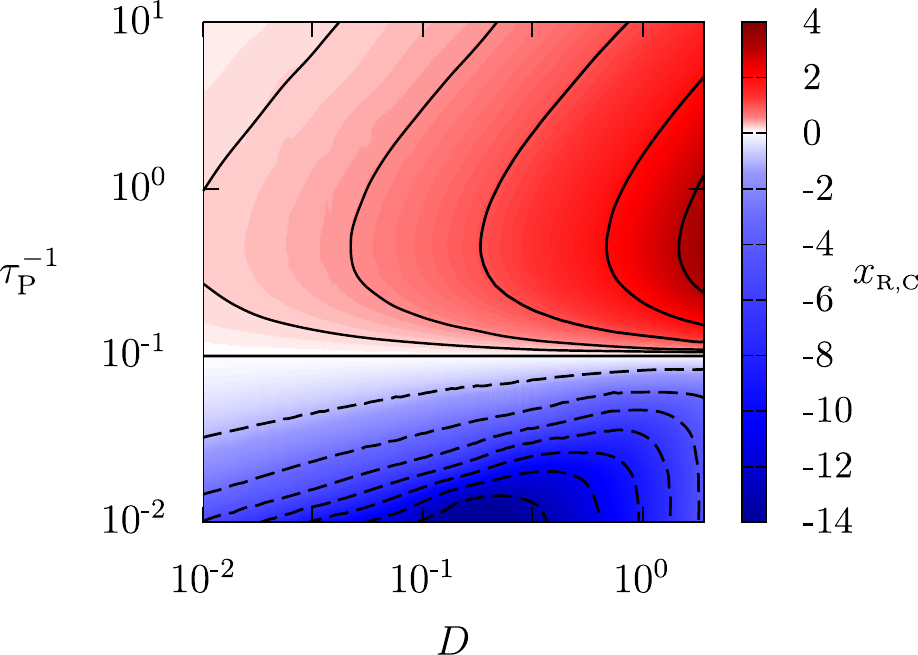}
    \caption{Critical location of the right habitat edge $x_{\s{R,C}}$ as a function of $\tau_{\s{P}}^{-1}$ and $D$ obtained through the semi-analytical method. Parameter values: $r_{\s{H}}=0.1$, $\gamma=1$, $x_{\s{L}}=-20$, $r_{\s{M}}\rightarrow-\infty$.}
	\label{fig:cps}
\end{figure}

For $\tau_{\s{P}}^{-1}>0.1$, movement bias is so strong that the attractive location must be within the habitat patch to avoid individuals entering the matrix and dying at a rate that cannot be outbalanced by population growth within the habitat (red region in Fig.\,\ref{fig:cps} and Fig.\,\ref{fig:SM2}). Moreover, due to strong bias toward the attractive location, the population is concentrated around that location. Increasing the diffusion coefficient $D$ makes $x_{\s{R,C}}$ increase because the population spreads out, and individuals become more likely to reach the matrix and die. Increasing $\tau_{\s{P}}^{-1}$ from $\tau_{\s{P}}^{-1}=0.1$ for a fixed diffusion coefficient $D$, we find a non-monotonic relationship between $x_{\s{R,C}}$ and $\tau_{\s{P}}^{-1}$. First, the critical patch size increases with $\tau_{\s{P}}^{-1}$ because the population concentrates around the attractive location and are more likely to reach the matrix and die. For even higher attraction rate, the population concentrates very narrowly around the attractive location and individuals do not reach the habitat edge. As a result, the critical patch size decreases with $\tau_{\s{P}}^{-1}$.

For $\tau_{\s{P}}^{-1}<0.1$, $x_{\s{R,C}}$ is negative, which means that the population can persist even when the attractive location is in the matrix (blue region in Fig.\,\ref{fig:cps}). In this low-$\tau_{\s{P}}^{-1}$ regime, $x_{\s {R,C}}$ increases with $\tau_{\s{P}}^{-1}$ because less random movement increases the relative contribution of the movement bias to the population flux through the edge. However, when $D$ increases, the critical patch size decreases, provided that $\tau_{\s{P}}^{-1}$ is not too far from $\tau_{\s{P}}^{-1}=0.1$. This negative correlation between critical patch size and diffusion appears because, \textit{when the attractive location is sufficiently close to a habitat edge}, a more random movement reduces the general tendency of individuals to leave the habitat through that edge. Fig.\,\ref{fig:SM2}b shows curves of $x_{\s{R,C}}$ versus $D$ with fixed $\tau_{\s{P}}^{-1}$. This phenomenon, known as the "drift paradox," has been previously observed in organisms inhabiting streams, rivers, and estuaries where downstream drift is continuously present, and extinction is inevitable in the absence of diffusion \citep{Pachepsky2005,speirs2001population}.
Nonetheless, as $D$ continues to increase and random diffusion dominates dispersal, the critical patch size increases due to population loss via diffusion through both habitat edges. Finally, for very low values of $\tau_{\s{P}}^{-1}$, diffusion controls the population flux through habitat edges and the behavior of the critical patch size converges to the theoretical prediction of the purely diffusive case, $L_{\s{C}}^{\s{D}}=\pi\sqrt{D/r_{\s{H}}}$ \citep{kierstead1953size}.

\subsection{Partially absorbing matrix and the effect of matrix-to-habitat bias}
\begin{figure}[!ht]
	\centering
	\includegraphics[width=0.8\columnwidth]{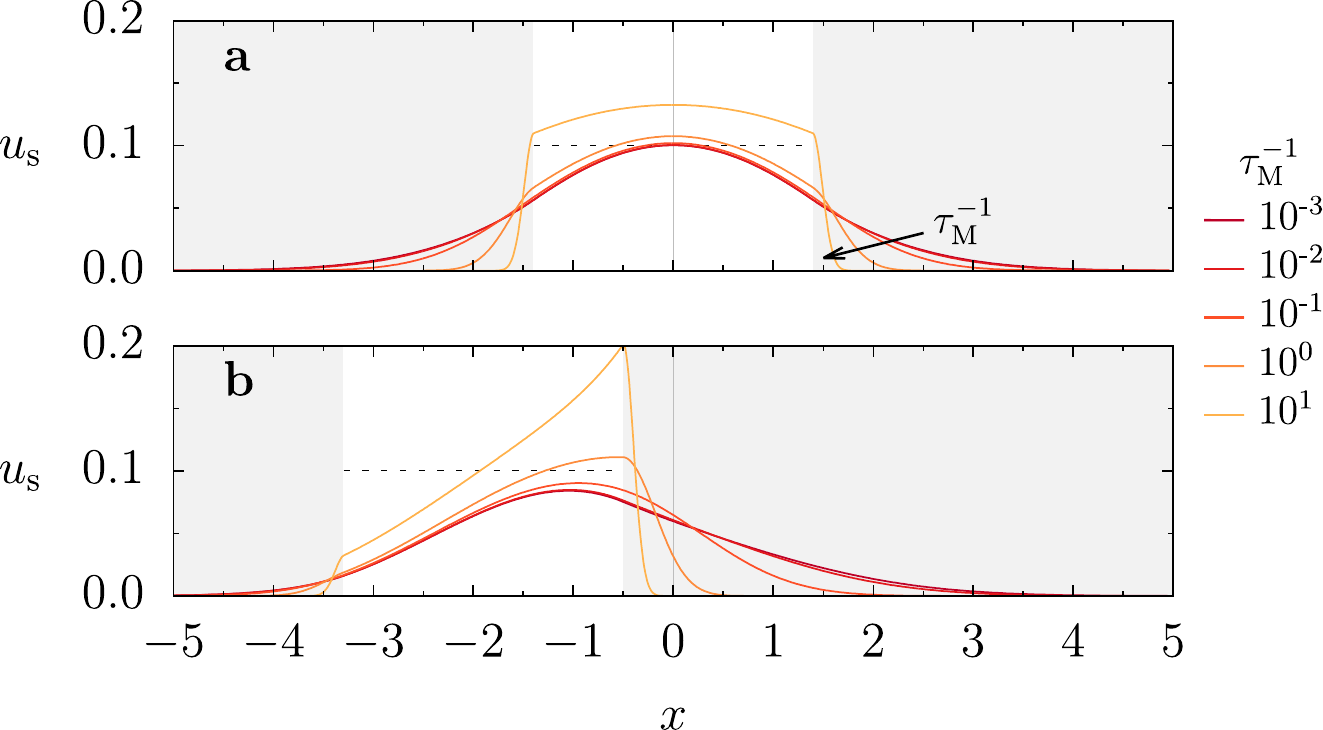}
	\caption{Stationary population density distribution $u_{\s{s}}(x)$ for the scenario with matrix-to-habitat attraction and attraction to the preferred location $x=x_{\s{P}}=0$ obtained through numerical simulation of Eq.\,\eqref{eq:PDE-popdens}. The parameters are $L=2.8$, which is smaller than the $L_{\s{C}}$ for $r_{\s{M}}\rightarrow-\infty$, $\tau_{\s{P}}^{-1}=0.05$, $r_{\s{M}}=-10^{-3}$, and  $\tau^{-1}_{\s{M}}$ increasing for lighter colors, as indicated. (a) $x_{\s{R}}=-x_{\s{L}}=1.4 $ and (b) $x_{\s{R}}=-0.5$ and $x_{\s{L}}=-3.3$. Other parameters as in Fig.\,\ref{fig:sppatt}. In both panels, the dashed line corresponds to $u_{\s{s}}=r/\gamma$, which is the density inside the habitat for $\tau^{-1}_{\s{M}}\rightarrow\infty$ and $\tau^{-1}_{\s{P}}=0$, in which case $u_{\s{s}}=0$ in the matrix.}
	\label{fig:uss-gm}
\end{figure}

Considering finite $r_{\s{M}}$ allows us to investigate how changes in movement behavior, once individuals reach the matrix, can alter the spatial pattern of population density and the critical patch size. If individuals in the matrix do not tend to return to the habitat ($\tau^{-1}_{\s{M}}\approx0$), the population density decays into the matrix exponentially, and the critical patch size increases with matrix mortality rate \citetext{Fig.\,\ref{fig:uss-gm}; \citealt{ludwig1979spatial}, \citealt{ryabov2008population}}. %However, the qualitative behavior of the critical patch size as a function of the movement parameters and the location of the habitat edges does not change with respect to the results for infinite matrix mortality rate presented in \ref{subsec:rminf}. 

For low values of $\tau^{-1}_{\s{M}}$, the tendency to return from the matrix to the habitat edges reduces how much the population penetrates the matrix and increases the population density inside the habitat, especially close to the edges (Fig.\,\ref{fig:uss-gm}a). The spatial distribution of the population has a skewness that reaches its maximum when the attractive location is in the matrix (Fig.\,\ref{fig:uss-gm}b). For large enough $\tau^{-1}_{\s{M}}$, we observe that the edges act as almost hard walls. This would be equivalent to having reflecting boundary conditions. In this limit, the population survives for any habitat size \citep{maciel2013individual}. 
%Moreover, if the attractive location is in the matrix, the density of individuals is maximum near the habitat edge. This accumulation of population density close to one of the edges is due to the balance between the bias toward the attractive location and the tendency to return from the matrix to the habitat (Fig.\,\ref{fig:uss-gm}b). For individuals located between the habitat edge and the attractive location, these two biases point in different directions.

The accumulation of individuals around habitat edges suggests a potential tradeoff between a decrease in mortality in the matrix due to the attraction to habitat edges and an increase in intraspecific competition due to higher population densities in the habitat. To investigate the impact of this tradeoff on population loss due to habitat degradation, we measure the fraction of the population that remains for a given patch size relative to the value for an infinite habitat patch, $\eta$. We perform this measurement for several values of the matrix mortality rate $r_{\s{M}}$ and the returning rate to habitat edges $\tau^{-1}_{\s{M}}$, which are the two main parameters controlling the accumulation of population density at habitat edges. We consider a scenario with the attractive location at the center of the habitat patch, which is the limit where we have a weaker accumulation of individuals at habitat edges and, therefore, the regime in which the tradeoff between matrix mortality and intraspecific competition around habitat edges has a weaker effect on population dynamics.

\begin{figure}[H]
	\centering
 \includegraphics[width=0.65\columnwidth]{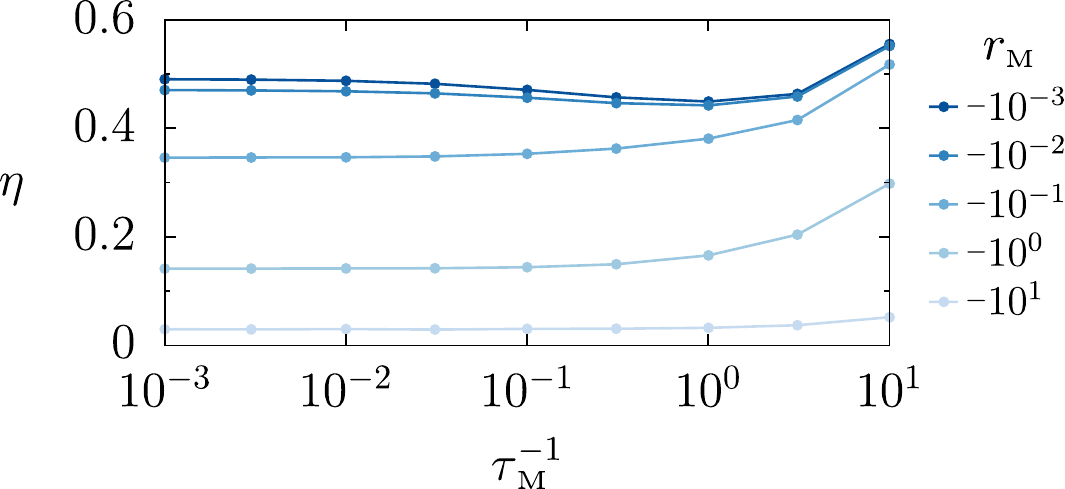}
	\caption{ Ratio between the total population for $L$ and for $L\rightarrow\infty$, $\eta$, versus $\tau^{-1}_{\s{M}}$ obtained through numerical simulations of Eq.\,\eqref{eq:PDE-popdens}. $x_{\s{R}}=-x_{\s{L}}=1.4 $ and values of $r_{\s{M}}$ as indicated, with lighter colors corresponding to larger absolute values of $r_{\s{M}}$. Other parameters as in Fig.\,\ref{fig:uss-gm}.}
	\label{fig:gm-eta}
\end{figure}

At high matrix mortality rates, the population does not survive ($\eta=0$), except for very high returning rates $\tau^{-1}_{\s{M}}$ (Fig.\,\ref{fig:gm-eta}). When the matrix mortality rate decreases, $\eta$ increases and remains a monotonically increasing function of $\tau^{-1}_{\s{M}}$. For $r_{\s{M}}$ closer to zero, however,
$\eta$ becomes a non-monotonic function of $\tau^{-1}_{\s{M}}$. %we observe a collapse of this monotonic trend between $\eta$ and $\tau^{-1}_{\s{M}}$. 
For these values of the matrix mortality rate, increasing the returning rate to habitat edges is initially detrimental to the total population size because it leads to higher intraspecific competition at the habitat edges, which outweighs the decrease in mortality in the matrix. In other words, the density distribution does not penetrate the matrix as far  (Fig.\,\ref{fig:uss-gm}a) while, inside the habitat, competition does not allow for a large enough increase in population, and so the total population decreases. Consequently, the habitat edge itself behaves as an ecological trap in this regime, and our model recovers a behavior similar to previous observations for insects \citep{ries2003habitat,ries2004ecological}. Above a critical value of $\tau^{-1}_{\s{M}}$ at which $\eta$ is minimal, further increasing the returning rate to habitat edges becomes beneficial for population persistence because now very few individuals enter the matrix and reduced matrix mortality outweighs the increased intraspecific competition at habitat edges. For infinite return rate $\tau^{-1}_{\s{M}}$, all the curves for different values of the matrix mortality rate $r_{\s{M}}$ converge to the same value because individuals do not penetrate the matrix. For $\tau^{-1}_{\s{M}}\rightarrow\infty$ and $\tau^{-1}_{\s{P}}=0$, one has $u_{\s{s}}=r/\gamma$ inside the habitat and $u_{\s{s}}=0$ in the matrix (dashed line in Fig.\,\ref{fig:uss-gm}). The existence of a non-monotonic dependence of population size on advection strength is reminiscent of a behavior reported in a different scenario for a model with advection towards a continuous environmental gradient \citep{belgacem1995effects}.

%=================================================
\section{Discussion}

We studied the spatial dynamics of a population in a finite habitat surrounded by an infinite matrix, considering different ratios between matrix mortality and habitat reproduction rates. We additionally incorporated space-dependent deterministic movement through an advection term that attracts individuals toward specific landscape locations, including habitat edges. This advection term can create spatial distributions of population density that are asymmetric with respect to the center of the patch, especially when the patch size is small and attractive regions lie near habitat edges. This result could explain why, in certain species, populations tend to accumulate in the periphery of the species historical range following geographical range contraction \citep{channell_trajectories_2000,channell2000dynamic}. Moreover, our results show that both the habitat carrying capacity and critical size depend nonlinearly, sometimes non-monotonically, on movement and demographic parameters and the location of the habitat edges relative to regions of slower movement. Recent work has also found nonlinear and non-monotonic relationships between movement and landscape parameters underlying the stability of prey-predator systems in fragmented landscapes \citep{Dannemann2018,Nauta2022}. These findings emphasize the importance of untangling the various contributions determining individual movement, including environmental covariates, when designing conservation strategies such as refuges in fragmented landscapes or marine protected areas \citep{gerber2003,gaston2002}. 

Specifically, for very low yet non-zero bias intensities, we find a range of values for the diffusion coefficient for which the critical habitat size decreases with increasing diffusion. In this parameter regime, if the attractive location is sufficiently close to a habitat edge, a more random movement reduces the general tendency of individuals to leave the habitat through that edge. This counterintuitive phenomenon is known as the ``drift paradox'' \citep{Pachepsky2005}. On the opposite limit, if movement bias toward the attractive location is very strong, the population becomes ultra-localized and its survival depends on whether the attractive site is in the habitat patch or the matrix; if it is in the patch, the population will persist, but if it is in the matrix, the population will go extinct. In between these two limits, for weak bias toward the attractive location, further increasing bias intensity increases the critical habitat size when the attractive site is inside the habitat but not too far from both edges. Moreover, populations are still viable for these weak bias intensities even if habitat destruction places the attractive location inside the matrix, creating an ecological trap. Ecological traps are often related to human landscape interventions \citep{schlaepfer2002,robertson2006} such as the construction of bird nest cavities in regions with generally worse conditions than those where the birds would naturally build their nests  \citep{krams2021ecological}. Roads can also act as ecological traps. For example, female bears with their cubs are often attracted to roads due to higher forage availability and to avoid potential male infanticide, increasing their risk of being killed in vehicle collisions \citep{penteriani2018,northrup2012}.

Our model also suggests that movement responses to changes in habitat quality, such as the tendency of individuals to return from the matrix to habitat edges, can result in the accumulation of population density around habitat edges, even when attractive locations are centered in the habitat patch. This accumulation of population density reduces the quality of regions nearby habitat edges relative to the surrounding matrix and turn the neighborhoods of habitat edges into ecological traps. This population crowding nearby habitat edges could, however, be eliminated by density-dependent dispersal, which was not included in the our model. Animal responses to changes in habitat fragmentation, such as the matrix avoidance term included in our model, might be relevant in regulating demographic responses to habitat destruction. Quantifying correlations between movement behavior, habitat quality, and population density in animal tracking data could help to understand the impact of further habitat destruction on population viability. More generally, the emergence of ecological traps (of the type induced by habitat loss) in our model suggests that movement patterns exhibited by individuals upon habitat destruction do not correspond to an evolutionarily stable strategy \citep{hastings1983can}. However, because ecological traps do not necessarily lead to population extinctions in our model, individuals could potentially adapt their movement behavior to avoid newly degraded regions. %Selection for these adapted movement strategies within the population could, on the long term, take the population to an evolutionarily stable dispersal strategy \citep{stephen2007ideal,maciel2020evolutionarily}.}

Different non-uniform space utilization patterns and preference for specific habitat locations are ubiquitous in nature. We consider that all individuals in the population have the same movement behavior and thus share habitat preferences. This assumption is an accurate modeling choice for certain species, such as central-place foragers \citep{fagan2007population}. Very often, however, habitat preferences vary across individuals in a population, which might impact how individuals interact with one another \citep{martinez2020range,noonan2021estimating}. Incorporating individual-level variability in space utilization would inform how populations of range-resident and territorial species would respond to habitat destruction, and is one of the future directions that could be explored based on this work. However, while attractiveness can sometimes be quantified in terms of environmental covariates \citep{mueller2008search} or by knowing the locations of landscape features like watering holes, other times it will be difficult or impossible to quantify, for example when ``attractiveness'' depends on the unknown distribution of a particular prey species. %We also neglected other features of animal movement that might have significant consequences for population-level responses to habitat destruction. For example, many organisms often exhibit highly autocorrelated velocities, even in homogeneous environments, which is important in determining space occupation patterns \citep{li2011, Noonan2019, villa2020run,reynolds2014} and the intensity of ecological interactions \citep{Noonan2023,Kiorboe2005}. As we showed with the simplest case of a space-dependent bias toward specific landscape locations, neglecting well-known properties of animal movement can result in inaccurate estimations of population-level responses to habitat fragmentation. 
Future theoretical research should aim to increasingly fill this gap between existing models describing empirically observed patterns of animal movement and higher level ecological processes.

\section*{Acknowledgments} We thank Silas Poloni, Eduardo H.~Colombo, and Chris Cosner for their critical reading of the manuscript. This work was partially funded by the Center of Advanced Systems Understanding (CASUS), which is financed by Germany’s Federal Ministry of Education and Research (BMBF) and by the Saxon Ministry for Science, Culture and Tourism (SMWK) with tax funds on the basis of the budget approved by the Saxon State Parliament; the S\~ao Paulo Research Foundation (FAPESP, Brazil) through postdoctoral fellowships No.~2021/10139-2 and No.~2022/13872-5 (P.d.C) and No.~2020/04751-4 (V.D.), BIOTA Young Investigator Research Grant No.~2019/05523-8 (R.M-G); ICTP-SAIFR grant no.~2021/14335-0 (P.d.C) and No.~2016/01343-7 (V.D., P.d.C., R.M-G); the Simons Foundation through grant no.\ 284558FY19 (R.M-G); the Estonian Research Council through grant PRG1059 (V.D.). The National Science Foundation (NSF, USA) grant DBI1915347 supported the involvement of J.M.C.~and W.F.F. This research was supported by resources supplied by the Center for Scientific Computing (NCC/GridUNESP) of the S\~ao Paulo State University (UNESP).

%\appendix

%\renewcommand\thefigure{S\arabic{figure}}   

%\bibliographystyle{elsarticle-harv} 
\bibliography{references}

\begin{thebibliography}{}

\bibitem[Alharbi and Petrovskii, 2016]{Alharbi2016}
Alharbi, W.~G. and Petrovskii, S.~V. (2016).
\newblock {The Impact of Fragmented Habitat's Size and Shape on Populations
  with Allee Effect}.
\newblock {\em Mathematical Modelling of Natural Phenomena}, 11(4):5--15.

\bibitem[Arfken and Weber, 1999]{arfken1999mathematical}
Arfken, G.~B. and Weber, H.~J. (1999).
\newblock Mathematical methods for physicists.

\bibitem[Ballard et~al., 2004]{Ballard2004}
Ballard, M., Kenkre, V.~M., and Kuperman, M.~N. (2004).
\newblock {Periodically varying externally imposed environmental effects on
  population dynamics}.
\newblock {\em Physical Review E}, 70(3):7.

\bibitem[Belgacem and Cosner, 1995]{belgacem1995effects}
Belgacem, F. and Cosner, C. (1995).
\newblock The effects of dispersal along environmental gradients on the
  dynamics of populations in heterogeneous environment.
\newblock {\em Canad. Appl. Math. Quart}, 3(4):379--397.

\bibitem[Cantrell and Cosner, 1991]{Cantrell1991}
Cantrell, R.~S. and Cosner, C. (1991).
\newblock {Diffusive Logistic Equations with Indefinite Weights: Population
  Models in Disrupted Environments II}.
\newblock {\em SIAM Journal on Mathematical Analysis}, 22(4):1043--1064.

\bibitem[Cantrell and Cosner, 1999]{Cantrell1999}
Cantrell, R.~S. and Cosner, C. (1999).
\newblock {Diffusion models for population dynamics incorporating individual
  behavior at boundaries: Applications to refuge design}.
\newblock {\em Theoretical Population Biology}, 55(2):189--207.

\bibitem[Cantrell and Cosner, 2001]{Cantrell2001}
Cantrell, R.~S. and Cosner, C. (2001).
\newblock {Spatial heterogeneity and critical patch size: Area effects via
  diffusion in closed environments}.
\newblock {\em Journal of Theoretical Biology}, 209(2):161--171.

\bibitem[Cantrell et~al., 2006]{cantrell2006movement}
Cantrell, R.~S., Cosner, C., and Lou, Y. (2006).
\newblock Movement toward better environments and the evolution of rapid
  diffusion.
\newblock {\em Mathematical biosciences}, 204(2):199--214.

\bibitem[Channell and Lomolino, 2000a]{channell2000dynamic}
Channell, R. and Lomolino, M.~V. (2000a).
\newblock Dynamic biogeography and conservation of endangered species.
\newblock {\em Nature}, 403(6765):84--86.

\bibitem[Channell and Lomolino, 2000b]{channell_trajectories_2000}
Channell, R. and Lomolino, M.~V. (2000b).
\newblock Trajectories to extinction: {Spatial} dynamics of the contraction of
  geographical ranges.
\newblock {\em Journal of Biogeography}, 27(1):169--179.

\bibitem[Colombo and Anteneodo, 2018]{colombo2018nonlinear}
Colombo, E. and Anteneodo, C. (2018).
\newblock Nonlinear population dynamics in a bounded habitat.
\newblock {\em Journal of Theoretical Biology}, 446:11--18.

\bibitem[Colombo and Anteneodo, 2016]{Colombo2016}
Colombo, E.~H. and Anteneodo, C. (2016).
\newblock {Population dynamics in an intermittent refuge}.
\newblock {\em Physical Review E}, 94(4):1--7.

\bibitem[Cronin et~al., 2020]{cronin2020modeling}
Cronin, J.~T., Fonseka, N., Goddard, J., Leonard, J., and Shivaji, R. (2020).
\newblock Modeling the effects of density dependent emigration, weak allee
  effects, and matrix hostility on patch-level population persistence.
\newblock {\em Mathematical Biosciences and Engineering}, 17(2):1718.

\bibitem[Cronin et~al., 2019]{cronin2019effects}
Cronin, J.~T., Goddard, J., and Shivaji, R. (2019).
\newblock Effects of patch--matrix composition and individual movement response
  on population persistence at the patch level.
\newblock {\em Bulletin of Mathematical Biology}, 81:3933--3975.

\bibitem[Dannemann et~al., 2018]{Dannemann2018}
Dannemann, T., Boyer, D., and Miramontes, O. (2018).
\newblock {L{\'{e}}vy flight movements prevent extinctions and maximize
  population abundances in fragile Lotka–Volterra systems}.
\newblock {\em Proceedings of the National Academy of Sciences},
  115(15):3794--3799.

\bibitem[Dornelas, 2023]{github}
Dornelas, V. (2023).
\newblock Spatial population modeling.
\newblock \url{https://github.com/VivianDornelas/SpatialPopulationModeling}.

\bibitem[Dos~Santos et~al., 2020]{dos2020critical}
Dos~Santos, M., Dornelas, V., Colombo, E., and Anteneodo, C. (2020).
\newblock Critical patch size reduction by heterogeneous diffusion.
\newblock {\em Physical Review E}, 102(4):042139.

\bibitem[Dunn and Gipson, 1977]{Dunn1977}
Dunn, J.~E. and Gipson, P.~S. (1977).
\newblock {Analysis of Radio Telemetry Data in Studies of Home Range}.
\newblock {\em Biometrics}, 33(1):85.

\bibitem[Fagan et~al., 1999]{Fagan1999}
Fagan, W.~F., Cantrell, R.~S., and Cosner, C. (1999).
\newblock {How habitat edges change species interactions}.
\newblock {\em American Naturalist}, 153(2):165--182.

\bibitem[Fagan et~al., 2009]{fagan2009interspecific}
Fagan, W.~F., Cantrell, R.~S., Cosner, C., and Ramakrishnan, S. (2009).
\newblock Interspecific variation in critical patch size and gap-crossing
  ability as determinants of geographic range size distributions.
\newblock {\em The American Naturalist}, 173(3):363--375.

\bibitem[Fagan et~al., 2007]{fagan2007population}
Fagan, W.~F., Lutscher, F., and Schneider, K. (2007).
\newblock Population and community consequences of spatial subsidies derived
  from central-place foraging.
\newblock {\em The American Naturalist}, 170(6):902--915.

\bibitem[Ferraz et~al., 2003]{ferraz2003rates}
Ferraz, G., Russell, G.~J., Stouffer, P.~C., Bierregaard~Jr, R.~O., Pimm,
  S.~L., and Lovejoy, T.~E. (2003).
\newblock Rates of species loss from amazonian forest fragments.
\newblock {\em Proceedings of the National Academy of Sciences},
  100(24):14069--14073.

\bibitem[Fleming et~al., 2014]{fleming2014fine}
Fleming, C.~H., Calabrese, J.~M., Mueller, T., Olson, K.~A., Leimgruber, P.,
  and Fagan, W.~F. (2014).
\newblock From fine-scale foraging to home ranges: a semivariance approach to
  identifying movement modes across spatiotemporal scales.
\newblock {\em The American Naturalist}, 183(5):E154--E167.

\bibitem[Franklin et~al., 2002]{Franklin2002}
Franklin, A.~B., Noon, B.~R., and George, T.~L. (2002).
\newblock What is habitat fragmentation?
\newblock {\em Studies in avian biology}, 25:20--29.

\bibitem[Gaston et~al., 2002]{gaston2002}
Gaston, K.~J., Pressey, R.~L., and Margules, C.~R. (2002).
\newblock Persistence and vulnerability: {Retaining} biodiversity in the
  landscape and in protected areas.
\newblock {\em Journal of Biosciences}, 27(4):361--384.

\bibitem[Gerber et~al., 2003]{gerber2003}
Gerber, L.~R., Botsford, L.~W., Hastings, A., Possingham, H.~P., Gaines, S.~D.,
  Palumbi, S.~R., and Andelman, S. (2003).
\newblock Population models for marine reserve design: {A} retrospective and
  prospective synthesis.
\newblock {\em Ecological Applications}, 13(sp1):47--64.

\bibitem[Hastings, 1983]{hastings1983can}
Hastings, A. (1983).
\newblock Can spatial variation alone lead to selection for dispersal?
\newblock {\em Theoretical Population Biology}, 24(3):244--251.

\bibitem[Holmes et~al., 1994]{Holmes1994}
Holmes, E.~E., Lewis, M.~A., Banks, J.~E., and Veit, R.~R. (1994).
\newblock Partial differential equations in ecology: Spatial interactions and
  population dynamics.
\newblock {\em Ecology}, 75(1):17--29.

\bibitem[Ibagon et~al., 2022]{Ibagon2022}
Ibagon, I., Furlan, A., and Dickman, R. (2022).
\newblock Reducing species extinction by connecting fragmented habitats:
  Insights from the contact process.
\newblock {\em Physica A: Statistical Mechanics and its Applications},
  603:127614.

\bibitem[Jeltsch et~al., 2013]{Jeltsch2013}
Jeltsch, F., Bonte, D., Pe'er, G., Reineking, B., Leimgruber, P., Balkenhol,
  N., Schr{\"o}der, B., Buchmann, C.~M., Mueller, T., Blaum, N., et~al. (2013).
\newblock Integrating movement ecology with biodiversity research-exploring new
  avenues to address spatiotemporal biodiversity dynamics.
\newblock {\em Movement Ecology}, 1(1):1--13.

\bibitem[Kapfer et~al., 2010]{kapfer2010modeling}
Kapfer, J., Pekar, C., Reineke, D., Coggins, J., and Hay, R. (2010).
\newblock Modeling the relationship between habitat preferences and home-range
  size: a case study on a large mobile colubrid snake from north america.
\newblock {\em Journal of Zoology}, 282(1):13--20.

\bibitem[Kawasaki et~al., 2012]{kawasaki2012impact}
Kawasaki, K., Asano, K., and Shigesada, N. (2012).
\newblock Impact of directed movement on invasive spread in periodic patchy
  environments.
\newblock {\em Bulletin of mathematical biology}, 74:1448--1467.

\bibitem[Kenkre and Kumar, 2008]{Kenkre2008}
Kenkre, V.~M. and Kumar, N. (2008).
\newblock {Nonlinearity in bacterial population dynamics: Proposal for
  experiments for the observation of abrupt transitions in patches}.
\newblock {\em Proceedings of the National Academy of Sciences of the United
  States of America}, 105(48):18752--18757.

\bibitem[Kierstead and Slobodkin, 1953]{kierstead1953size}
Kierstead, H. and Slobodkin, L.~B. (1953).
\newblock The size of water masses containing plankton blooms.
\newblock {\em J. mar. Res}, 12(1):141--147.

\bibitem[Krams et~al., 2021]{krams2021ecological}
Krams, R., Krama, T., Br{\=u}melis, G., Elferts, D., Strode, L.,
  Dau{\v{s}}kane, I., Luoto, S., {\v{S}}mits, A., and Krams, I.~A. (2021).
\newblock Ecological traps: evidence of a fitness cost in a cavity-nesting
  bird.
\newblock {\em Oecologia}, 196(3):735--745.

\bibitem[Lamb et~al., 2017]{lamb2017forbidden}
Lamb, C.~T., Mowat, G., McLellan, B.~N., Nielsen, S.~E., and Boutin, S. (2017).
\newblock Forbidden fruit: human settlement and abundant fruit create an
  ecological trap for an apex omnivore.
\newblock {\em Journal of Animal Ecology}, 86(1):55--65.

\bibitem[Lin et~al., 2004]{Lin2004}
Lin, A.~L., Mann, B.~A., Torres-Oviedo, G., Lincoln, B., K{\"{a}}s, J., and
  Swinney, H.~L. (2004).
\newblock {Localization and extinction of bacterial populations under
  inhomogeneous growth conditions}.
\newblock {\em Biophysical Journal}, 87(1):75--80.

\bibitem[Lord and Norton, 1990]{Lord1999}
Lord, J.~M. and Norton, D.~A. (1990).
\newblock Scale and the spatial concept of fragmentation.
\newblock {\em Conservation Biology}, 4(2):197--202.

\bibitem[Ludwig et~al., 1979]{ludwig1979spatial}
Ludwig, D., Aronson, D., and Weinberger, H. (1979).
\newblock Spatial patterning of the spruce budworm.
\newblock {\em Journal of Mathematical Biology}, 8(3):217--258.

\bibitem[Maciel and Lutscher, 2013]{maciel2013individual}
Maciel, G.~A. and Lutscher, F. (2013).
\newblock How individual movement response to habitat edges affects population
  persistence and spatial spread.
\newblock {\em The American Naturalist}, 182(1):42--52.

\bibitem[Martinez-Garcia et~al., 2020]{martinez2020range}
Martinez-Garcia, R., Fleming, C.~H., Seppelt, R., Fagan, W.~F., and Calabrese,
  J.~M. (2020).
\newblock How range residency and long-range perception change encounter rates.
\newblock {\em Journal of theoretical biology}, 498:110267.

\bibitem[Mueller et~al., 2008]{mueller2008search}
Mueller, T., Olson, K.~A., Fuller, T.~K., Schaller, G.~B., Murray, M.~G., and
  Leimgruber, P. (2008).
\newblock In search of forage: predicting dynamic habitats of mongolian
  gazelles using satellite-based estimates of vegetation productivity.
\newblock {\em Journal of Applied Ecology}, 45(2):649--658.

\bibitem[Nathan et~al., 2008]{Nathan2008}
Nathan, R., Getz, W.~M., Revilla, E., Holyoak, M., Kadmon, R., Saltz, D., and
  Smouse, P.~E. (2008).
\newblock A movement ecology paradigm for unifying organismal movement
  research.
\newblock {\em Proceedings of the National Academy of Sciences},
  105(49):19052--19059.

\bibitem[Nauta et~al., 2022]{Nauta2022}
Nauta, J., Simoens, P., Khaluf, Y., and Martinez-Garcia, R. (2022).
\newblock {Foraging behavior and patch size distribution jointly determine
  population dynamics in fragmented landscapes}.
\newblock {\em Journal of The Royal Society Interface}, 19:20220103.

\bibitem[Noonan et~al., 2021]{noonan2021estimating}
Noonan, M.~J., Martinez-Garcia, R., Davis, G.~H., Crofoot, M.~C., Kays, R.,
  Hirsch, B.~T., Caillaud, D., Payne, E., Sih, A., Sinn, D.~L., et~al. (2021).
\newblock Estimating encounter location distributions from animal tracking
  data.
\newblock {\em Methods in Ecology and Evolution}, 12(7):1158--1173.

\bibitem[Noonan et~al., 2019]{Noonan2019}
Noonan, M.~J., Tucker, M.~A., Fleming, C.~H., Akre, T.~S., Alberts, S.~C., Ali,
  A.~H., Altmann, J., Antunes, P.~C., Belant, J.~L., Beyer, D., Blaum, N.,
  B{\"{o}}hning-Gaese, K., Cullen, L., {De Paula}, R.~C., Dekker, J.,
  Drescher-Lehman, J., Farwig, N., Fichtel, C., Fischer, C., Ford, A.~T.,
  Goheen, J.~R., Janssen, R., Jeltsch, F., Kauffman, M., Kappeler, P.~M., Koch,
  F., LaPoint, S., Markham, A.~C., Medici, E.~P., Morato, R.~G., Nathan, R.,
  Oliveira-Santos, L. G.~R., Olson, K.~A., Patterson, B.~D., Paviolo, A.,
  Ramalho, E.~E., R{\"{o}}sner, S., Schabo, D.~G., Selva, N., Sergiel, A.,
  {Xavier da Silva}, M., Spiegel, O., Thompson, P., Ullmann, W., Zi{\c{e}}ba,
  F., Zwijacz-Kozica, T., Fagan, W.~F., Mueller, T., and Calabrese, J.~M.
  (2019).
\newblock {A comprehensive analysis of autocorrelation and bias in home range
  estimation}.
\newblock {\em Ecological Monographs}, 89(2):e01344.

\bibitem[Northrup et~al., 2012]{northrup2012}
Northrup, J.~M., Pitt, J., Muhly, T.~B., Stenhouse, G.~B., Musiani, M., and
  Boyce, M.~S. (2012).
\newblock Vehicle traffic shapes grizzly bear behaviour on a multiple-use
  landscape.
\newblock {\em Journal of Applied Ecology}, 49(5):1159--1167.

\bibitem[Okubo, 1972]{okubo1972note}
Okubo, A. (1972).
\newblock A note on small organism diffusion around an attractive center: A
  mathematical model.
\newblock {\em Journal of the Oceanographical Society of Japan}, 28(1):1--7.

\bibitem[Pachepsky et~al., 2005]{Pachepsky2005}
Pachepsky, E., Lutscher, F., Nisbet, R.~M., and Lewis, M.~A. (2005).
\newblock {Persistence, spread and the drift paradox}.
\newblock {\em Theoretical Population Biology}, 67(1):61--73.

\bibitem[Penteriani et~al., 2018]{penteriani2018}
Penteriani, V., Delgado, M. D.~M., Krofel, M., Jerina, K., Ordiz, A., Dalerum,
  F., Zarzo-Arias, A., and Bombieri, G. (2018).
\newblock Evolutionary and ecological traps for brown bears \textit{{Ursus}
  arctos} in human-modified landscapes.
\newblock {\em Mammal Review}, 48(3):180--193.

\bibitem[Pereira and Daily, 2006]{pereira2006}
Pereira, H.~M. and Daily, G.~C. (2006).
\newblock Modeling biodiversity dynamics in countryside landscapes.
\newblock {\em Ecology}, 87(8):1877--1885.

\bibitem[Perry, 2005]{Perry2005}
Perry, N. (2005).
\newblock {Experimental validation of a critical domain size in
  reaction-diffusion systems with Escherichia coli populations}.
\newblock {\em Journal of the Royal Society Interface}, 2(4):379--387.

\bibitem[Press, 2007]{press2007numerical}
Press, W.~H. (2007).
\newblock {\em Numerical recipes 3rd edition: The art of scientific computing}.
\newblock Cambridge university press.

\bibitem[Ries and Fagan, 2003]{ries2003habitat}
Ries, L. and Fagan, W.~F. (2003).
\newblock Habitat edges as a potential ecological trap for an insect predator.
\newblock {\em Ecological entomology}, 28(5):567--572.

\bibitem[Ries et~al., 2004]{ries2004ecological}
Ries, L., Fletcher~Jr, R.~J., Battin, J., and Sisk, T.~D. (2004).
\newblock Ecological responses to habitat edges: mechanisms, models, and
  variability explained.
\newblock {\em Annu. Rev. Ecol. Evol. Syst.}, 35:491--522.

\bibitem[Robertson and Hutto, 2006a]{robertson2006framework}
Robertson, B.~A. and Hutto, R.~L. (2006a).
\newblock A framework for understanding ecological traps and an evaluation of
  existing evidence.
\newblock {\em Ecology}, 87(5):1075--1085.

\bibitem[Robertson and Hutto, 2006b]{robertson2006}
Robertson, B.~A. and Hutto, R.~L. (2006b).
\newblock A framework for understanding ecological traps and an evaluation of
  existing evidence.
\newblock {\em Ecology}, 87(5):1075--1085.

\bibitem[Ryabov and Blasius, 2008]{ryabov2008population}
Ryabov, A.~B. and Blasius, B. (2008).
\newblock Population growth and persistence in a heterogeneous environment: the
  role of diffusion and advection.
\newblock {\em Mathematical Modelling of Natural Phenomena}, 3(3):42--86.

\bibitem[Schlaepfer et~al., 2002]{schlaepfer2002}
Schlaepfer, M.~A., Runge, M.~C., and Sherman, P.~W. (2002).
\newblock Ecological and evolutionary traps.
\newblock {\em Trends in ecology \& evolution}, 17(10):474--480.

\bibitem[Skellam, 1951]{Skellam1951}
Skellam, J. (1951).
\newblock {Random Dispersal in Theoretical Populations}.
\newblock {\em Biometrika}, 38(1):196--218.

\bibitem[Speirs and Gurney, 2001]{speirs2001population}
Speirs, D.~C. and Gurney, W.~S. (2001).
\newblock Population persistence in rivers and estuaries.
\newblock {\em Ecology}, 82(5):1219--1237.

\bibitem[Turner, 1996]{turner1996species}
Turner, I.~M. (1996).
\newblock Species loss in fragments of tropical rain forest: a review of the
  evidence.
\newblock {\em Journal of applied Ecology}, pages 200--209.

\bibitem[Tyson et~al., 2007]{tyson2007modelling}
Tyson, R., Thistlewood, H., and Judd, G.~J. (2007).
\newblock Modelling dispersal of sterile male codling moths, cydia pomonella,
  across orchard boundaries.
\newblock {\em Ecological modelling}, 205(1-2):1--12.

\bibitem[Van~Moorter et~al., 2016]{van2016movement}
Van~Moorter, B., Rolandsen, C.~M., Basille, M., and Gaillard, J.-M. (2016).
\newblock Movement is the glue connecting home ranges and habitat selection.
\newblock {\em Journal of Animal Ecology}, 85(1):21--31.

\bibitem[Vergni et~al., 2012]{Vergni2012}
Vergni, D., Iannaccone, S., Berti, S., and Cencini, M. (2012).
\newblock {Invasions in heterogeneous habitats in the presence of advection}.
\newblock {\em Journal of Theoretical Biology}, 301:141--152.

\bibitem[Weldon and Haddad, 2005]{weldon2005effects}
Weldon, A.~J. and Haddad, N.~M. (2005).
\newblock The effects of patch shape on indigo buntings: evidence for an
  ecological trap.
\newblock {\em Ecology}, 86(6):1422--1431.

\bibitem[Zhou and Fagan, 2017]{zhou2017discrete}
Zhou, Y. and Fagan, W.~F. (2017).
\newblock A discrete-time model for population persistence in habitats with
  time-varying sizes.
\newblock {\em Journal of Mathematical Biology}, 75:649--704.

\end{thebibliography}





% \end{thebibliography}

\newpage
\appendix

\renewcommand\thefigure{S\arabic{figure}}    

\section*{Supplementary figures}
\setcounter{figure}{0}      

\begin{figure}[H]
    \centering
    \includegraphics[width=0.6\columnwidth]{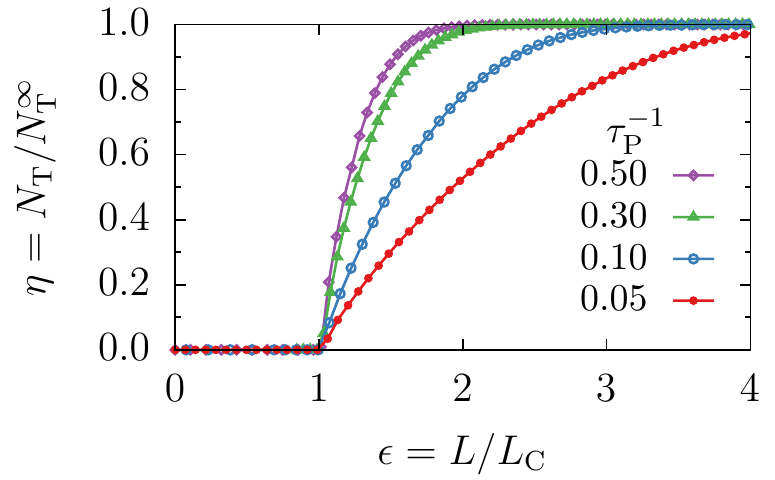}
    \caption{Ratio between the total population for $L$ and for $L\rightarrow\infty$, $\eta$, versus $\epsilon\equiv L/L_{\rm C}$ for various $\tau_{\s{P}}^{-1}$ as indicated obtained through numerical simulations of Eq.\,\eqref{eq:PDE-popdens}. Other parameters: $D=0.1$, $\gamma=1$, $r_{\rm H}=0.1 $, and  $r_{\s{M}}\rightarrow-\infty$. The habitat is symmetric, i.e., $x_{\rm R}=-x_{\rm L}$.}
    \label{fig:SM1}
\end{figure}

\begin{figure}[H]
    \centering	\includegraphics[width=0.9\columnwidth]{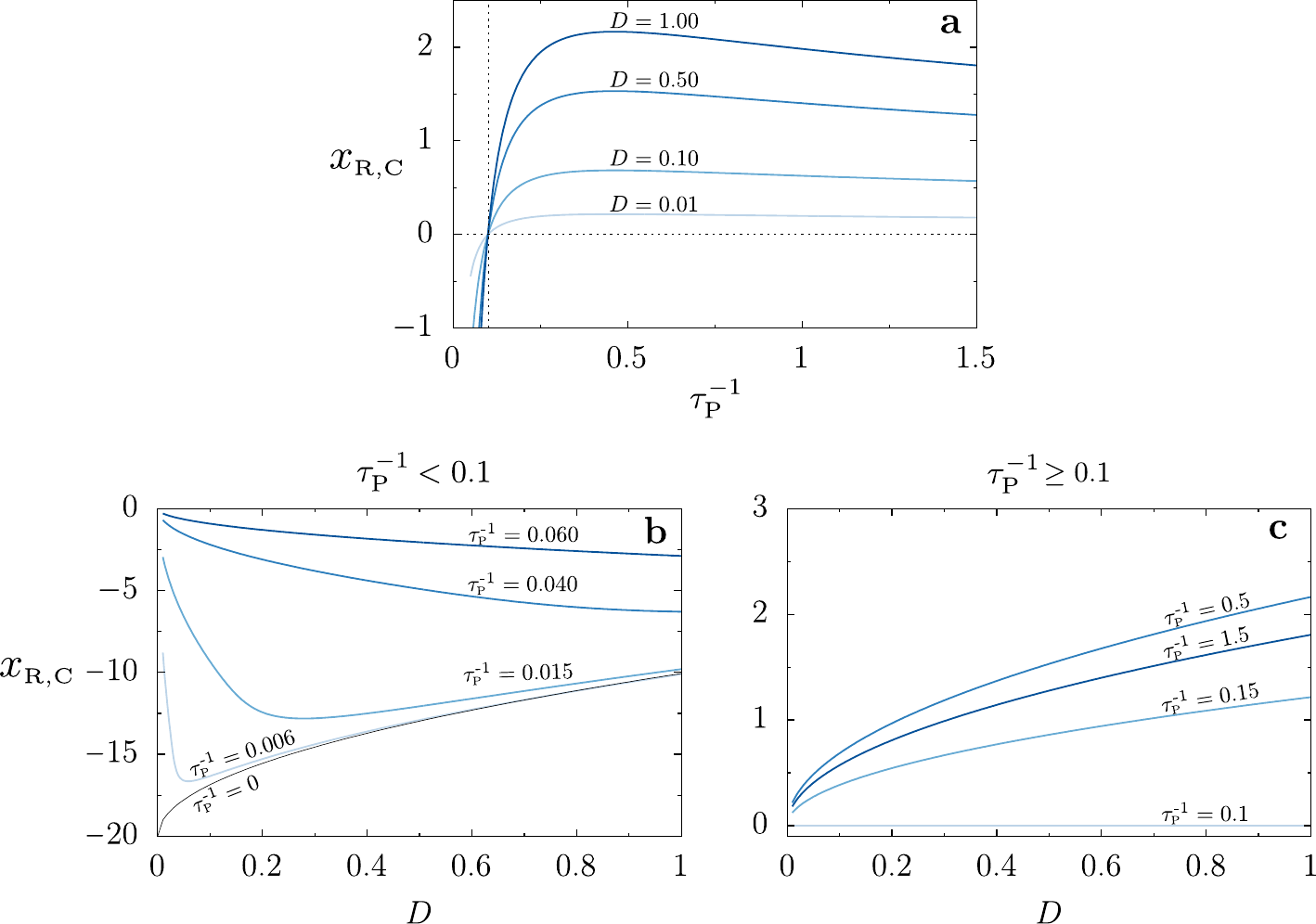}
    \caption{Critical location of the right habitat edge $x_{\s{R,C}}$ (a) as a function of $\tau_{\s{P}}^{-1}$, and as a function of $D$ for (b) $\tau_{\s{P}}^{-1}<0.1$ and (c) $\tau_{\s{P}}^{-1}>0.1$, obtained through the semi-analytical method. Other parameters: $r_{\s{H}}=0.1$, $\gamma=1$, $r_{\s{M}}\rightarrow-\infty$, and $x_{\s{L}}=-20$.}
    \label{fig:SM2}
\end{figure}

\end{document}